\documentclass[%
article,%
 amssymb, amsmath,%
 aip,cha,%
]{revtex4-1}

\usepackage{graphicx}
\usepackage{txfonts}
\usepackage{docs}%
\usepackage{bm}%
\usepackage[colorlinks=true,linkcolor=blue]{hyperref}%
\expandafter\ifx\csname package@font\endcsname\relax\else
 \expandafter\expandafter
 \expandafter\usepackage
 \expandafter\expandafter
 \expandafter{\csname package@font\endcsname}%
\fi
\begin{document}

\title{Derivation of the Planck Mass from Gravitational Polarization of the Quantum Vacuum}%

\author{Martin Tajmar}%
\email{martin.tajmar@tu-dresden.de}
\affiliation{Instiute for Aerospace Engineering, Dresden University of Technology, 01062 Dresden, Germany}%

\date{March 2012}%
%\revised{August 2010}%

\begin{abstract}
The Planck units were originally derived from a dimensional analysis without a deeper understanding of their meaning. It was later believed that these units may provide a link between quantum theory and gravity in a yet to be developed theory of quantum gravity. I propose a model where the Planck units appear naturally by assuming the the quantum vacuum can be gravitationally polarized based on recent work on the gravitational properties of anti-particles. In order to match the observed values, we arrive at Planck particle/anti-particle pairs (micro black holes) with Planck masses that define the gravitational constant in vacuum through gravitational polarization. This gives the Planck mass a new important interpretation as indeed linking quantum fluctuations to gravity by defining the gravitational constant. In addition, a better understanding of why the Planck length is usually associated as the smallest length in nature can be illustrated from another perspective.

\keywords{Planck Units, Vacuum Polarization, Gravitational Schwinger Limit}

\end{abstract}

\maketitle

\section{Introduction}
In his 1899 paper and 1906 book about heat radiation \cite{planck1899,planck1906}, Planck defined a natural scale from dimensional analysis for time, length and mass composed of physical constants involving only the speed of light $c$, the gravitational constant $G$ and the (later defined as) reduced Planck constant $\hbar$ given by

\begin{equation}
m_P=\sqrt{\frac{\hbar c}{G}}, l_P=\sqrt{\frac{\hbar G}{c^3}}, t_P=\sqrt{\frac{\hbar G}{c^5}} 
\end{equation}

Later, the Planck mass $m_P$ was given a physical meaning as being the mass of a tiny black hole, which has a Schwarzschild radius approximately equal to its Compton wavelength. The Planck units are believed to be crucial for a quantum theory of gravity, but apart from a dimensional analysis, a deeper understanding of this link from these units is missing and a successful quantum gravity theory is still to be developed \cite{weinstein2011}.

As I will show in this paper, the Planck mass and length appear naturally when the gravitational permittivity and permeability are computed using the latest models of quantum vacuum fluctuations. This gives the Planck units a deeper physical interpretation beyond the original dimensional analysis. It may help to see how gravity and the quantum world may be connected at these scales.

\section{Gravitational Polarization of the Quantum Vacuum} 
In quantum field theory, the vacuum is filled with virtual particle-antiparticle pairs. For example, the electron-positron pairs create electric dipoles, which are polarized in the presence of an electric field. This is called vacuum polarization and plays an important role in charge renormalization. A number of related effects are predicted such as vacuum birefringence using intense magnetic or electric fields (Schwinger critical field) which is presently under investigation using laser beams (e.g. \cite{heinzl2006, dellavalle2010}). 

Recently, Hajdukovic \cite{hajdukovic2010,hajdukovic2010EPL,hajdukovic2011} proposed that the quantum vacuum also contains virtual gravitational dipoles on which an external gravitational field can act. He assumes that the gravitational mass of anti-particles are negative ($\overline{m}_g=-m_g$) and that the inertial mass of anti-particles, as known by observation, is positive ($\overline{m}_i=m_i$). Negative anti-particle gravitational masses are not ruled out by observation so far and dedicated experiments are planned to investigate this issue (e.g. the AEGIS experiment at CERN \cite{kellerbauer2008}). In another recept paper, Villata \cite{villata2011} concludes that negative gravitational masses appear as a prediction of general relativity when CPT is applied. Of course, negative gravitational masses would be the most drastic violation of the weak-equivalence principle possible - but as long as no gravitational experiments with anti-particles have been carried out, it is a theoretical possibility. The properties of anti-gravitational matter is discussed at length in the references give above.

Hajdukovic \cite{hajdukovic2010,hajdukovic2010EPL,hajdukovic2011} further assumed, that only virtual pions are important for the gravitational interaction with the vacuum. In his simplified analysis, he assumed a fully gravitational polarized vacuum to explain the Pioneer anomaly as well as dark energy. By calculating the gravitational permittivity of vacuum from the hypothesis of virtual gravitational dipoles only, I can show that not virtual pions but rather virtual Planck particles (with Planck mass) lead to the correct gravitational behaviour. For completeness, we should note that the concept of gravitational vacuum polarization and associated shielding phenomenon were discussed in the literature back in the 1970s and 1980s most notably by Long \cite{long1980} and others (e.g. \cite{shao1989}).

We revisit this subject without any prior assumption of a particular virtual particle or sum of particles and only assume the existance of gravitational dipoles in the quantum vacuum. Then we follow a similar approach as Leuchs et al \cite{leuchs2010}, who were able to derive the electric permittivity of vacuum using the polarization of virtual electric dipoles from the vacuum (a similar derivation was recently published by Urban et al \cite{urban2011}). However, they modeled the virtual electric dipole as a harmonic oscillator in the quasi-static limit that is possible for electrically attracting dipoles but not a priori for gravitationally repulsing gravitational dipoles. We will propose a modified model assuming a negative spring constant.

In our case, the virtual pair starts from the same position and the particles repel each other. The characteristic time that they travel is given by the Heisenberg uncertainty principle and is given by

\begin{equation}
\Delta t \cong \frac{\hbar}{2 \Delta E}=\frac{1}{2\omega_0} ,
\end{equation}

where $E=\hbar \omega_0=2mc^2$ is the rest energy of the virtual dipole and $\omega_0$ is the fundamental resonance frequency associated with the quantum transition. Instead of a sinusoidal solution for the harmonic oscillator, the position of a mass connected to a negative spring constant is given as

\begin{equation}
x(t)=x_0 \cdot e^{-\omega t}(e^{2 \omega_0 t}-1) ,
\end{equation}

where we have assumed that the pair starts a position zero at $t=0$. The acceleration is then given by

\begin{equation}
\ddot{x}(t)=x_0 \omega_0^2 \cdot e^{-\omega_0 t}(e^{2 \omega_0 t}-1) ,
\end{equation}

Using a Taylor series expansion for the exponential function and setting $t=\Delta t$, we can simplify this expression assuming a small $\Delta t$ to

\begin{equation}
\ddot{x}(t) \cong 2x_0 \omega_0^2 (\omega_0 \Delta t - \omega_0^2 \Delta t^2) \cong 2x_0 \omega_0^3 \Delta t= x_0 \omega_0^2
\end{equation}

It is interesting to note that this result is similar to the harmonic oscillator quasi-static limit as assumed by Leuchs et al 
\cite{leuchs2010}. The force balance under the influence of an external gravitational field $g$ is now given as

\begin{equation}
m x_0 \omega_0^2 = m g
\label{distance}
\end{equation}

We continue by defining the gravitational dipole moment as $p_g=mx_0$, where $m$ is the mass of the virtual particle and $x_0$ the displacement that we just derived. Now the gravitational dipole moment is given as

\begin{equation}
p_g=\frac{m}{\omega^2_0}g 
\end{equation}

The density of dipoles is usually expressed by the fact that a virtual dipole approximately occupies the space given by the reduced Compton wavelength of the virtual particles such that $N=1/ \lambdabar^3$ with $\lambdabar=\hbar / m c$. We can now express the gravitational polarization of the vacuum $P_g=N.p_g$ as

\begin{equation}
P_g=N \cdot \frac{m}{\omega^2_0}g = \frac{m}{\omega^2_0\lambdabar^3}g
\label{polarization}
\end{equation}

The induced polarization is proportional to the gravitational field $g$ and hence we can assign this proportionality factor to the gravitational permittivity of vacuum $P_g=\epsilon_g.g$ (similarily to the electric case, where the polarization is the product of the electric permittivity with the applied electric field $P_0=\epsilon_o . E$). We can express it as

\begin{equation}
\epsilon_g=\frac{m}{\omega^2_0\lambdabar^3}=\frac{m^2}{4\hbar c}
\label{epsilon_g}
\end{equation}

Leuchs et al \cite{leuchs2010} noted in their analysis for the electric case, that such an expression must not be due to the contribution from only one particle, and hence the virtual particle mass $m$ may not be the mass of a single particle but the contribution from (in our case) all virtual particle masses. Leuchs et al estimated between 10-90 charged particles contributing to the electric permittivity by comparing their result from a single electron-positron dipole to the known electric constant. For the gravitational polarization contribution, it would be natural to assume that the virtual dipole masses add up to a maximum value limited by their Compton wavelength and the Schwarzschild radius, which of course is the definition of the Planck mass. We will see that this is indeed the case by equalling the gravitational permittivity from our virtual dipole fluctuations in Equ. \ref{epsilon_g} to the well known permittivity from the Einstein-Maxwell equations, which are derived by liniarizing general relativity in the weak-field approximation \cite{tajmar2001,kiefer2005}. We then get

\begin{equation}
\epsilon_g=\frac{m^2}{4\hbar c} = \frac{1}{4 \pi G}
\label{epsilon_gravitational_constant}
\end{equation}

Solving for the mass of our virtual gravitational dipoles, we get

\begin{equation}
m=\sqrt{\frac{\hbar c}{G \pi}} = \frac{m_P}{\sqrt{\pi}}
\end{equation}

The hypothesis of the existance of gravitational dipoles leads us to the conclusion that the strength of the gravitational interaction in vacuum (=gravitational constant) can be expressed as a gravitational polarization of the vacuum from dipoles with Planck masses. The gravitational permittivity can now also be expressed in Planck units as

\begin{equation}
\epsilon_g=\frac{m_P}{\pi \omega^2_{0P} l_P^3}=\frac{m_P^2}{4\pi\hbar c} = \frac{1}{4 \pi G}
\end{equation} 

Our model is of course only valid for distances larger than the Planck length $l_P$. However, there is an interesting difference between the electric and gravitational vacuum polarization as we can point out as follows. In the electric case, vacuum polarization does not play a role at distances larger than the electron Compton wavelength and at field strengths below the Schwinger limit. However, when approaching this distance, electric vacuum polarization becomes noticable as a shielding effect (since non-equal charges attract) and the Coulomb law needs to be modified as first calculated by Uehling \cite{uehling1935} and Serber \cite{serber1935}. So in fact, the real electric charge value is larger than it seems at macroscopic distances. For the gravitational case, we also expect a modification from Newton's law when we approach the Planck length but with a different sign resulting in larger than real gravitational field since in this case equal gravitational charges attract. Therefore, the real masses are in fact smaller as obtained by measuring at macroscopic distances.

A fascinating question, of course, is what happens to gravity below the Planck length? Although our model assumes distances larger than $l_P$, we can point out another interesting difference compared to the electric case. In Leuchs's model \cite{leuchs2010}, the density of virtual dipoles is directly proportional to the electric permittivity/constant $\epsilon_0 \propto N$. For a scale smaller than the electron-positron pair wavelength, the density of virtual dipoles $N=0$ and hence the vacuum contribution inside that scale is zero and only the outside dipoles contribute (i.e. the Uehling potential modification to Coulomb's law \cite{uehling1935}). For gravity, the situation is different. Here, the virtual dipole density is indirectly proportional to the gravitational constant $G \propto 1/N$ (see Equs. \ref{polarization} and \ref{epsilon_gravitational_constant}). When the scale is smaller than $l_P$ and $N=0$, the model predicts, that the vacuum contribution inside that scale now is not zero, as in the electric case, but infinite. We may interpret this difference that the Planck length is the smallest length scale in nature bacause an infinite gravitational constant reduces any mass distribution inside that scale to a point.

We can also approach this from another point of view. In the electromagnetic case, the Schwinger limit is the electric field $E=\frac{m^2m^3}{eh}\cong 10^{18} V/m$ at which the distance between virtual electric dipoles approaches the Compton wavelength. Beyond that length, the virtual particles become real. We can define a similar case for gravity. When the gravitational field become so large that the distance between the fluctuating dipoles approaches the Compton wavelength, we reach a Gravitational Schwinger Limit. The critical gravitational field may be calculated by rearranging Equ.~\ref{distance} as

\begin{equation}
x_0=\frac{mg_{critical}}{m \omega^2} \cong \lambdabar
\end{equation}

and we get with using the Planck mass for the fluctuating particles

\begin{equation}
g_{critical}=\frac{mc^3}{\hbar}=\frac{c^4}{G}=10^{44} \frac{m}{s^2}
\end{equation}

This is similar to the electromagnetic Schwinger limit if we replace the charge with the mass. This critical field is reached at the Schwarzschild radius (=Planck length) of the Planck particle. At distances below the Planck length, the virtual Planck masses (micro Black-holes) would become real. This may also support the few that the Planck length is the smallest length scale in nature protected by a gravitational type Schwinger limit.

\section{Conclusion}
I proposed a model that can derive the Planck units out of the existence of recently proposed fluctuating gravitational dipoles as part of the quantum vacuum. In order to match the observed gravitational constant, the model predicts that Planck particle/anti-particle pairs (micro black holes) exist in the vacuum, which are polarized by a gravitational field, giving the Planck mass an important interpretation beyond just originating from a dimensional analysis. 

The model can help to answer also questions such as why the Planck length is usually associated with the smallest length scale in all modern physical theories. Also a gravitational-type Schwinger limit was derived that again illustrates how gravity can become important at small scales.

The Planck mass of the virtual dipoles shows, that not only virtual pions are responsible for gravitational interaction as assumed by Hajdukovic \cite{hajdukovic2010,hajdukovic2010EPL,hajdukovic2011}. They may be considered in addition to the Planck particles to account for MOND or dark energy effects. Our Planck mass model predicts, that we are surrounded by large amounts of anti-matter with negative gravitational masses, which may lead to observable effects that should be further explored.

%\section{References}

\bibliographystyle{unsrt}
\bibliography{Bibliography}

\end{document}